\documentclass[%
reprint,
nofootinbib,
amsmath,amssymb,
aps,
floatfix
]{revtex4-2}

\usepackage{graphicx}
\usepackage{dcolumn}
\usepackage{bm}
\usepackage{hyperref}

\usepackage{color}
\usepackage[dvipsnames]{xcolor}

\hypersetup{colorlinks=true, citecolor=MidnightBlue,linkcolor=CornflowerBlue, urlcolor=CornflowerBlue, linktocpage=true}

\usepackage[section]{placeins}

\usepackage{placeins}

\newcommand\Lie{\pounds}

\newcommand{\beq}{\begin{eqnarray}}
\newcommand{\eeq}{\end{eqnarray}}
\newcommand{\beqn}{\begin{eqnarray}}
\newcommand{\eeqn}{\end{eqnarray}}
\newcommand{\pa}{\partial}
\newcommand{\rd}{\mathrm{d}}
\newcommand{\un}[1]{\underline{#1}}

\begin{document}

\title{Fluid-gravity correspondence and causal first-order 
relativistic viscous hydrodynamics}

\author{Luca Ciambelli}
 \email{ciambelli.luca@gmail.com}
\author{Luis Lehner}%
\email{llehner@perimeterinstitute.ca}
\affiliation{%
 Perimeter Institute for Theoretical Physics, \\
 31 Caroline St North, Waterloo ON N2L 2Y5, Canada
}

\date{\today}


\vspace{1cm}

\begin{abstract}
The fluid-gravity correspondence is a duality between anti-de Sitter Einstein gravity and a relativistic fluid living at the conformal boundary. We show that one can accommodate the causal first-order viscous hydrodynamics recently developed by Bemfica, Disconzi, Noronha, and Kovtun in this framework, by requiring a set of natural conditions for the geometric data at the horizon. The latter hosts an induced Carrollian fluid, whose equations of motion are shown to be tightly tied to the ones describing the fluid at the boundary. Functional expressions for the transport coefficients are found --with those
associated to viscosity and heat flux uniquely determined--, 
satisfying a set of known causality requirements for the underlying equations
of motion. 
\end{abstract}

\maketitle

\section{Introduction} \label{sec:Introduction}
In a series of works by Bemfica, Disconzi, Noronha, and Kovtun (BDNK), a formulation for viscous, relativistic hydrodynamics
has been introduced where dissipative corrections are accounted for via 
first-order derivatives of the energy density and flow velocity \cite{Bemfica:2017wps,Kovtun:2019hdm,Bemfica:2019knx}, see also \cite{Hoult:2020eho, Bemfica:2020zjp, Hoult:2021gnb},
and where causality of the resulting equations of motion is achieved
when choosing transport coefficients within particular bounds.
Such a formulation, at first sight, is in tension with standard results where
at least second-order corrections are required to account for viscous relativistic 
hydrodynamics~\cite{GEROCH1991394,Denicol:2008ha,Pu:2009fj,Lehner:2017yes,Aguilar:2017ios}. A key observation is that such
results require a strictly non-negative entropy change, while the
first-order formulation does so up to higher order in gradients. Arguably, this is not 
necessarily a significant shortcoming
as such  higher order terms should be subleading
in the effective field theory regime where such theory can be
written. Further, another key aspect 
of BDNK is formulating such theory in a more 
general frame than the typical Landau or Eckart frames where
causality is violated in the first order formulation. 

The Landau frame \cite{1959flme} was introduced requiring that the heat current vanishes, such that the fluid velocity is an eigenvector of the energy-momentum tensor. On the other hand, in the Eckart frame \cite{Eckart:1940te} the fluid velocity is aligned with the  particle number flux, such that the equations are similar to those of non-relativistic hydrodynamics. As pointed out by BDNK, the frame discussed above should not be chosen driven by aesthetics, but instead requiring that the resulting hydrodynamic equations lead to well-posed problems, thus the equations of motions should be hyperbolic and causal.

In a parallel development, the celebrated fluid-gravity correspondence~\cite{Bhattacharyya:2007vjd, Bhattacharyya:2008xc,VanRaamsdonk:2008fp, Haack:2008cp, Banerjee:2008th, Rangamani:2009xk}
has linked the behavior of perturbed black holes (with asymptotically  anti-de Sitter boundary
conditions) to viscous relativistic hydrodynamics in one lower dimension. This remarkable
correspondence, was fully developed to second order in gradients, but 
specialized in the Landau frame by judicious choices made when solving Einstein equations 
for a perturbed black brane. Under restricted assumptions on the bulk duals, the Landau frame was abandoned in \cite{Caldarelli:2011idw, Mukhopadhyay:2013gja, Gath:2015nxa}, where the heat current was considered in the fluid-gravity correspondence.

In the current work, with the aim of shedding light on the connection between
the fluid-gravity correspondence and the 
BDNK first order, viscous relativistic hydrodynamics, 
we first show that the fluid-gravity correspondence is well-suited to accommodate the full first order hydrodynamic frame spectrum. This freedom was already present in \cite{Bhattacharyya:2007vjd}, but the
correspondence was fully developed in the Landau frame. Given this freedom, are there reasonable choices --in particular, from a gravity perspective-- leading to BDNK? 
To answer this question, we study the properties of the bulk projected on the horizon, which is a null hypersurface. 

It is known since the original ``membrane paradigm'' \cite{Damour:1979wya, Thorne:1986iy} that the Einstein equations projected to the horizon are conservation equations of a fluid, which has been recently understood to be a Carrollian fluid \cite{Penna:2018gfx,Donnay:2019jiz,Ciambelli:2019lap,Freidel:2022bai, Freidel:2022vjq,Redondo-Yuste:2022czg,Ciambelli:2023mir}. We show that the Carrollian equations of motion, at first order, are equal to those of
perfect fluid conservation for a conformal fluid\footnote{In particular,
in agreement with the boundary fluid equations of motion for the perfect fluid.}. We also observe that requiring the null vector generating the horizon to be aligned with the fluid velocity at first order in the derivative expansion selects exactly the BDNK form of the heat current. Similarly, the energy density at first order is the one used by BDNK if it is proportional to the horizon expansion. Under these assumptions we derive the values induced by Einstein gravity of most transport coefficients and a condition on a remaining one for a conformal equation of state. We find that the transport coefficients than can be fully fixed this way are within the causality and stability bounds. These  observations may open the door to 
unraveling a deeper connection between the horizon Carrollian fluid and  relativistic conformal fluids. 

The rest of the manuscript is organized as follows. In section \ref{sec:setup} we review the construction  of the fluid-gravity correspondence of \cite{Bhattacharyya:2007vjd}, and extrapolate the boundary energy-momentum tensor using the holographic prescription \cite{Henningson:1998gx, Balasubramanian:1999re}. In section \ref{sec:choices} we discuss the horizon geometry using the framework of \cite{Mars:1993mj}, and study it at zeroth and first order derivative expansion. We then make our geometrical choices, that we show in section \ref{sec:consequences} lead to BDNK for the boundary fluid. We conclude with final remarks in section \ref{sec:final}.

\section{Setup} \label{sec:setup}
Following the presentation in~\cite{Bhattacharyya:2007vjd}, 
we consider a boosted black brane in 4+1 dimensions
with asymptotically anti-de Sitter boundary conditions.
In the stationary case --zeroth order in the gradient expansion--, the spacetime metric is given by
\beqn
\rd s^2 = - 2 u_a \rd x^a \rd r+r^2(P_{ab}-f(br) u_a u_b) 
\rd x^a \rd x^b \, ;
\eeqn
with $f(r)=1-r^{-4}$, $u^a u^b \eta_{ab}=-1$ and $P_{ab} = \eta_{ab} + u_a u_b$ the projector orthogonal to $u^a$, $u^aP_{ab}=0$. The vector $u^a$ is constant and defines the boost, while the function $f(br)$ describes a black
brane with radius $r_H=b^{-1}$. Perturbed solutions, in terms of
a gradient expansion (also known as derivative expansion) can be obtained by considering $(b,u^a)$ as slowly varying functions of $x^a$\footnote{We use bulk coordinates $x^\mu=(r,x^a)$, such that $x^a$ are coordinates of fixed-$r$ hypersurfaces, and in particular of the boundary.}, inserting into Einstein equations
and solving for non-trivial functions of $r$ (see~~\cite{Bhattacharyya:2007vjd}).

As opposed to the treatment in that work, we refrain
from adopting the specific choice of vanishing ``zero-modes''
(as also considered in~\cite{Hoult:2021gnb}).
In other terms, we allow for small coordinate changes in the $x^a$ sector, 
which we capture in terms of a scalar function
$\chi(x^a)$ and a vector $q^a(x^b)$ transverse to $u^a$, $u^a q_a=0$.
The resulting solution, at first order in the gradient expansion is\footnote{Using the convention $A_{(ab)}=\frac12(A_{ab}+A_{ba})$.}
\beqn
\rd s^2 &=& - 2 u_a \rd x^a \rd r+r^2(P_{ab}-f(br) u_a u_b) \rd x^a \rd x^b\nonumber \\
& & -\frac2{b^4 r^2} u_{({a}} q_{{b})} \rd x^{a} \rd x^{b} +\frac{\chi}{b^4r^2}  u_{a} u_{b} \rd x^{a} \rd x^{b} \nonumber \\
& & + 2 r^2 b F(br) \sigma_{ab}\rd x^a \rd x^b
+\frac23 r \theta  u_{a} u_{b} \rd x^{a} \rd x^{b} \nonumber \\
&& -2r u_{({a}} a_{{b})} \rd x^{a} \rd x^{b},\label{fom}
\eeqn
where, asymptotically\footnote{$F(r)$ is given by 
a trascendental expression, see eqn. (4.20) in~\cite{Bhattacharyya:2007vjd} for
details.}, $F(br)=\frac1{br}-\frac1{4b^4r^4}$, and we introduced the shear, expansion, and acceleration of $u^a$,
\beq
&\sigma^{ab}=P^{c (a}\pa_c u^{b)}-\frac13 P^{ab}\pa_c u^c&\\
&a_a=u^b\pa_b u_a\, , \qquad \theta=\pa_a u^a,&
\eeq
satisfying $u^a\sigma_{ab}=0$, $\sigma_a{}^a=0$ ($\eta_{ab}P^{ab}=3$),  $u^a a_a=0$.

Armed with this solution, we can now obtain the stress
tensor induced at the timelike asymptotic boundary. To do so, we follow the holographic prescription discussed in \cite{Henningson:1998gx, Balasubramanian:1999re}, of which we now recap the salient ingredients. Given a bulk metric $g_{\mu\nu}$, we introduce a hypersurface at fixed $r$ and its projector $h_{\mu\nu}$. Using the normalized normal form $N=N_\mu \rd x^\mu=\frac{\rd r}{\sqrt{g^{rr}}}$, the projector reads
\beq
h_{\mu\nu}=g_{\mu\nu}-N_\mu N_\nu\quad h_\mu{}^\nu N_\nu=0.
\eeq
The extrinsic curvature (second fundamental form) is defined as
\beq
K_{ab}&=&h_a{}^\mu h_{b}{}^\nu\frac12 \Lie_N g_{\mu\nu}\\
&=&h_a{}^\mu h_{b}{}^\nu\frac12 (\nabla_\mu N_\nu+\nabla_\nu N_\mu),
\eeq
where $\nabla$ is the bulk Levi-Civita connection.
The induced inverse metric is
\beq
\overline{g}^{ab}=h_\mu{}^a h_\nu{}^b g^{\mu\nu},
\eeq
which can be used to define the trace on the hypersurface
\beq
K=\overline{g}^{ab}K_{ab}.
\eeq
The traceless part of the extrinsic curvature defines the boundary stress tensor
\beq
T^a{}_b=-2\lim_{r\to \infty}r^4\left(K^a{}_b-\frac{K}{4}\delta^a_b\right),
\eeq
where the $r$ pre-factor comes from the holographic dictionary and ensures its finiteness approaching the boundary.

Applying this procedure to our line element \eqref{fom}, the final result is 
\beq\label{bst}
T_{ab}=3\frac{1+\chi}{b^4} u_a u_b +\frac{1+\chi}{b^4}P_{ab}-\frac2{b^3}\sigma_{ab}-\frac{8}{b^4}u_{(a}q_{b)}.
\eeq
This is the stress tensor of a conformal viscous fluid with fluid velocity $u^a$, which 
accounts for heat-flux through $q^a$, and a correction to
the perfect fluid energy density through $\chi$.
Since a generic stress tensor is decomposed as
\beq\label{tab}
T_{ab}={\cal E} u_a u_b +{\cal P}P_{ab}-2\eta\sigma_{ab}+u_{a}Q_{b}+Q_{a}u_{b},
\eeq
one has,
\beqn \label{stresstensor_quants}
{\cal E}&=&3\frac{1+\chi}{b^4},\quad {\cal P}=\frac{1+\chi}{b^4}\quad  \nonumber \\
\eta&=&\frac1{b^3}\, , \quad Q_a=-\frac4{b^4}q_a.
\eeqn
We note that ${\cal E}=3{\cal P}$ is the conformal equation of state implied by the asymptotic conformal
symmetry, streaming from the vanishing of the stress tensor trace. Notice that at equilibrium the temperature is given by $T=\frac1b$.

From here, one could straightforwardly identify conditions for
$(\chi,q^a)$ to recover BDNK. This would amount to using a particular
formulation of viscous-relativistic hydrodynamics to fix 
conditions on the gravitational sector.  However, our goal is to
go in the opposite way, namely to consider arguably natural choices
on the gravitational sector --specifically at the horizon-- and explore what they correspond to in the
hydrodynamic side.

\section{Choices} \label{sec:choices}

To consistently deal with a degenerate metric at the null
hypersurface describing the horizon, we adopt the null Rigging formalism described
in~\cite{Mars:1993mj}. The horizon is generically located at $r=r_H(x)$, and thus the one-form normal to the horizon is
\beq
\un{n}=\tilde \alpha \rd( r-r_H(x)),
\eeq
and we adopt $\tilde \alpha=1$ in the following.
Next, we introduce the vector $k=\partial_r$ with the defining properties
\beq
\un n(k)=1\, , \qquad \un k(k)=0.
\eeq
This vector is called the {\em null Rigging vector}. We can then
define the Rigging projector as
\beq
\Pi_\mu{}^\nu=\delta_\mu^\nu-n_\mu k^\nu,
\eeq
such that
\beq
\Pi_\mu{}^\nu n_\nu=0\qquad k^\nu\Pi_\nu{}^\mu=0.
\eeq
The Rigging projector projects to the null hypersurface, since indeed the form $\un n$ and the vector $k$ are normal to it.

The bulk metric duals
$n$ and $\un k=-u_a \rd x^a$ satisfy
\beq
\Pi_\mu{}^\nu k_\nu=k_\mu\qquad \ell^\mu=n^\nu\Pi_\nu{}^\mu.
\eeq
Furthermore, the projected metric is given by
\beq
q_{\mu\nu}&=&\Pi_\mu{}^\rho\Pi_\nu{}^\sigma g_{\rho\sigma}=g_{\mu\nu}-n_\mu k_\nu-k_\mu n_\nu.
\eeq
The components intrinsic to the hypersurface, $(k_a,\ell^a,q_{ab})$, form the ruled Carrollian structure discussed in \cite{Ciambelli:2023mir} (with the same conventions). In particular, $\ell^a$ is the Carrollian vector field, $k_a$ is the Ehresmann connection, and $q_{ab}$ is the degenerate Carrollian metric satisfying
\beq
\ell^a q_{ab}=0
\eeq
at the horizon.

The other relevant quantities for the horizon physics are the surface gravity, expansion, H\'ajiček connection, and acceleration. 
They are defined in the bulk by, \\
(1) Surface gravity:\footnote{This quantity should be called inaffinity, but for non-expanding horizons these two concepts coincide. Here, by construction at zeroth order and as a consequence of the equations of motion at first order, the horizon expansion vanishes so we are in this framework.}
\beq
\ell^\mu\nabla_\mu \ell^\nu=\kappa \ell^\nu, \qquad 
 k_\nu \ell^\mu\nabla_\mu \ell^\nu=\kappa  \, ;
\eeq
(2) Expansion:
\beq
\Theta = q_\nu{}^\mu \nabla_\mu n^\nu \, ;
\eeq
(3) H\'ajiček connection: 
\beq
\pi_\mu=q_\mu{}^\nu k_\rho \nabla_\nu n^\rho \, ;
\eeq
(4) Acceleration 
\beq
\varphi_\mu=n^\nu\nabla_{[\mu}k_{\nu]} \, .
\eeq

We now proceed to compute these quantities, for clarity
we do so first in the stationary solution (zeroth order), 
and then in the first order perturbed case.

\subsection{Zeroth Order}

At this order, the location of the horizon, and associated
normal form and vector are: 
\beq
r_H=\frac1b \,, \qquad \un n=\rd r\, ,\qquad k=\pa_r \, .
\eeq
The bulk metric duals are
\beq
n=r^2 f(br)\pa_r+u^a\pa_a\, , \qquad \un k=-u_a \rd x^a
\eeq
and thus the Carrollian vector is exactly given by the boundary fluid congruence (which is constant at zeroth order)
\beq
\ell^\mu=n^\nu\Pi_\nu{}^\mu= u^a \delta_a^\mu.
\eeq
This implies
\beq
\ell^\mu k_\mu=-u^a u_a=1.
\eeq
The degenerate metric on the null surface is
\beq
q_{\mu\nu}=
\begin{pmatrix}
    0 & 0\\
    0 & r^2(P_{ab}-f(br) u_a u_b)
\end{pmatrix} \xrightarrow{r=r_H}\begin{pmatrix}
    0 & 0\\
    0 & \frac{P_{ab}}{b^2}
\end{pmatrix}
\eeq
which indeed satisfies at the horizon
\beq
\ell^\mu q_{\mu \nu}=u^a q_{ab}\delta^b_\nu\xrightarrow{r=r_H}u^a \frac{P_{ab}}{b^2}=0.
\eeq
With the above quantities, it is straightforward to obtain:
\beq
\kappa=\frac2b, \quad \Theta=0, \quad \pi_\mu=0, \quad \varphi_\mu=0 .
\eeq
Exactly like the relativistic conformal fluid at the boundary, the Carrollian fluid at the horizon is a perfect fluid at zeroth order. 
Delving into the properties of the Carrollian fluid on
the horizon and its connection to the boundary fluid would bring us too afield from the subject of this manuscript. We leave this 
exploration to a future work.

\subsection{First Order}

We now perturb the stationary solution using the first order gradient expansion. Details on how to establish the location of the perturbed horizon are in \cite{Bhattacharyya:2008xc} (in particular subsection 2.3), so we just report the result here.
At first order, the horizon and associated normal form are
\beq
r_H=\frac1b+\frac{\theta}{6}+\frac{\chi}{4b}-\frac{u^a\pa_a b}{2b}, \qquad \un n=\rd r+\frac{\rd b}{b^2},
\eeq
where $\theta$, $\chi$, and $\rd b$ are first order quantities.

Following the steps described above, we gather
\beq
&\un k=-u_a \rd x^a,&\\  &\ell^a=u^a-ba^a-q^a+P^{ab}\pa_b b\, , \quad \ell^r=-u^a \frac{\pa_a b}{b^2};&
\eeq
where the indices of the various quantities ($a^a$, $q^a$, and $P_{ab}$) are raised using $\eta_{ab}$, and we note that $\ell^r$ is non-vanishing due to the fact that the horizon position is now a function of $x^a$.

With the above, through a direct, but naturally more involved
calculation, one obtains:
\beqn
\kappa&=&\frac2b-\frac{2\theta}{3}+\frac{\chi}{2b}+\frac{u^a\pa_a b}{b}\label{kappa}\\
\Theta &=& \theta-3\frac{u^a\pa_a b}{b}\label{Th}\\
\pi_a&=&2\frac{q_a}{b}-\frac{P_a{}^b\pa_b b}{b}\\
\varphi_a&=&a_a.\label{phi}
\eeqn
With these, we are now ready to argue for some particular
choices. 
First, one could demand that at first order, the component of
the null vector $\ell^\mu$ orthogonal to $r$ should be aligned with $u^a$
(just as in the zeroth order case). This allows one to still identify the Carrollian vector with the boundary fluid velocity, even at first order. Such a choice implies
\beq
q^{a} = -b a^{a} + P^{ab} \partial_{b} b \, .
\eeq
This, as we shall discuss below, is precisely in line with the hydrodynamic
choice in BDNK. 

Before such discussion, we must address the
choice of $\chi$. First, note that $r_H$ can be re-expressed as,
\beq
r_H = \frac{1}{b} + \frac{1}{6} \Theta + \frac{\chi}{4b}.
\eeq
We shall now show that, to first order, $\Theta=0$ as a consequence of the Einstein equations projected on the horizon, specifically the Raychaudhuri equation. Thus, the choice
$\chi \propto \Theta$, conveniently keeps
$r_H = 1/b$ on-shell. Note that, with this choice,  
$\kappa$ receives non-trivial first-order corrections.
We discuss the consequences of choosing it to remain unchanged
in appendix~\ref{altchoice}. Since $\kappa$ depends on the generators' parameterization, we regard keeping $r_H$ unchanged as the more
natural choice.

To see that $\Theta=0$ to first order, let us recall that 
Raychaudhuri's and Damour's equations in vacuum are, 
\beqn \label{Ray}
&(\Lie_{\ell}+\Theta)[\Theta] = \mu \Theta -\sigma_{a}{}^b\sigma_b{}^a 
,&\\ \label{Damour}
&q_a{}^b \left(\Lie_{\ell}+\Theta\right)[\pi_b]+\Theta\varphi_a = (\overline{D}_b+\varphi_b)(\mu q_a{}^b-\sigma_{a}{}^b)&
\eeqn
where $\Lie_{\ell}$ is the Lie-derivative along $\ell$,
$\overline{D}_a$ the (Carrollian) covariant derivative associated to $q_{ab}$,
$\mu = \Theta/2 + \kappa$ and we used the conventions of \cite{Ciambelli:2023mir}. Since here we will be interested only in the first order expression, where most terms in these equations vanish, we refer to this reference for an explanation of all the quantities involved in general. Notice $\kappa$ has an order 0 contribution, so, eq. (\ref{Ray}) implies
that at first order $\Theta = 0$. A similar analysis of eq. (\ref{Damour}) implies $q_a{}^b\pa_b \kappa + \varphi_a \kappa =0$, where $q_a{}^b$ is the projector orthogonal to $\ell^a$ at the horizon, which at zeroth order is simply $P_a{}^b$, and thus (using \eqref{phi}) this equation is equal to $a_a=P_a{}^b\frac{\pa_b b}{b}$.

These observations have several consequences. First, since to the order we work, $\Theta=0$, the choice
stated above for $\chi$ indeed implies $r_H=1/b$ to this order. Further, and importantly, they indicate 
that at first order Raychaudhuri's and Damour's equations are exactly equal to the conservation of the boundary perfect fluid stress tensor. Indeed, one can easily show that $\pa_a T^a{}_b=0$, using \eqref{bst} and the relationships \eqref{kappa}, \eqref{Th}, and \eqref{phi}, gives exactly the Raychuadhuri's and Damour's equations, once projected on $u^a$ and $P_a{}^b$, respectively. This is ultimately tied to the fact that these equations all come from the bulk Einstein equations
and their particular hierarchical structure arising from the
characteristic treatment along a timelike-null foliation of
the spacetime.

To summarize then, examining the resulting structure at the horizon, our choices are:
\beqn
&\ell^a=u^a \Leftrightarrow   q^{a} = -b a^{a} + P^{ab} \partial_{b} b \, &\\
&r_H = \frac{1}{b} + \left(2+3\alpha\right) \frac{\Theta}{12} \Leftrightarrow  \chi = \alpha \, b\, \Theta \, ,&
\eeqn
with $\alpha$ a proportionality function that remains to be specified. We reported these results off-shell of the conservation laws discussed above. If we now impose these conservation laws, we obtain $q^a=0$ and $\chi=0$. This is precisely the outcome of the intrinsic hydrodynamic BDNK analysis for a conformal relativistic fluid: the heat current and the first-order correction to the energy that implement causality  are zero on-shell of the first order conservation law. In
what follows, we discuss in detail the structure implied by
the geometrical identifications/choices on the resulting
hydrodynamical equations.

\section{Consequences} \label{sec:consequences}

We can now examine the consequences of these choices on the thermodynamic quantities obtained
in (\ref{stresstensor_quants}). First, note that
\beqn
{\cal E}^{(0)}= 3{\cal P}^{(0)} &\equiv& e =\frac{3}{b^4} \label{energy0}\\
{\cal E}^{(1)}= 3{\cal P}^{(1)}&=&\frac{3 \alpha}{b^3} \left(\partial_{a} u^{a} - \frac{3}{b} 
u^{a} \partial_{a} b \right) \\ 
Q^{a} &=& \frac{4}{b^3} \left( a^{a} -  \frac{ P^{ab} \partial_{b} b}{b} \right),
\eeqn
where we introduced $\{e,p=\frac{e}{3}\}$ to denote the zeroth order expressions for
energy and pressure respectively.

The first order expressions
can be re-expressed in terms of $e$ and $p$ as,
\beqn
{\cal E}^{(1)}&=& \frac{3 \alpha}{b^3} \left(\partial_{a} u^{a} + \frac{u^{a} \partial_{a} e}{(e+p)} 
 \right) \\ 
Q^{a} &=& \frac{4}{b^3} \left( a^{a} +  \frac{ P^{ab} \partial_{b} e}{3 (e+p)} \right) 
\eeqn
(the expressions for the pressure is trivially set by the conformal
condition).
We can now compare with the expressions adopted by BDNK for
the conformal case, as this is the one that corresponds to
our case~\cite{Bemfica:2019knx}. Namely, denoting with an overbar
their choices,
\beqn
\bar {\cal E}^{(1)}&=&\left(\chi_2 \, \partial_{a} u^{a} + \chi_1 \,\frac{u^{a} \partial_{a} e}{(e+p)} 
 \right) \\ 
\bar Q^a&=&\lambda \left( a^{a} +  \frac{ P^{ab} \partial_{b} e}{3 (e+p)} \right),
\eeqn
with $\lambda,\chi_i$ transport coefficients\footnote{
Which include $\{\chi_3,\chi_4\}$ analogously introduced
for the first-order pressure ${\cal P}^{(1)}$.} that are chosen to ensure 
causality of the underlying equations, together with $\eta$ defined in \eqref{tab}.

Remarkably, the functional form
for the first order corrections are in excellent agreement
with the proposed terms in~\cite{Bemfica:2019knx}.
Moreover, our choices motivated by considerations at the horizon also imply for the transport coefficients (for $\eta$ we recall \eqref{stresstensor_quants}),
\beqn
&\eta = \frac{1}{b^3} \, , \, \lambda=\frac{4}{b^3} \, ,& \nonumber \\
&\chi_1 =\chi_2=3\chi_3=3\chi_4= \frac{3 \alpha}{b^3} \, ,&
\eeqn
where $\{\chi_3,\chi_4\}$ are linked to
 $\{\chi_1,\chi_2\}$ by conformality.

Not only do the transport coefficients have the temperature dependency of $T^{3}$ as expected from kinetic theory~\cite{Bemfica:2019knx}, but the shear viscosity and heat transport coefficients are uniquely determined\footnote{The value of the viscous transport coefficient is tied to the lowest-lying quasinormal modes of the perturbed black brane (see, e.g.~\cite{Berti:2009kk}).}.
In particular,
they satisfy the criteria for causality $\lambda \ge \eta$ identified in~\cite{Bemfica:2019knx}. Notice however our expressions make the transport coefficients $\chi_i$ all proportional to each other but do not 
completely fix them, nor provide bounds for them
which need not be surprising.  Namely,
conditions on $\chi_i$ determined by the causality 
analysis of~\cite{Bemfica:2019knx}, effectively, come 
from the high frequency limit (through the standard analysis
within PDE theory). 
This can be seen by examining the
dispersion relations for the shear and sound modes and their dependency
on $\{\eta,\lambda,\alpha\}$. Their roles appear at order $k^2$, $k^4$ and
$k^6$ respectively. On the other hand, the  fluid-gravity correspondence is obtained in the long wavelength regime of perturbed black holes
in General Relativity --which is a causal theory--, thus it is
natural to expect that in the regime where the duality
can be established, conditions on relevant parameters
on the hydrodynamic side can be obtained implying such property. 

For the unfixed parameter, we only demand $\alpha > 0$, as this choice ensures
equilibrium can be reached, i.e. ${\cal E}^{(0)} + {\cal E}^{(1)} \rightarrow {\cal E}^{(0)}$ within a timescale given by $\alpha$. Of course, one can choose
a suitable value for $\alpha$ such that the 
full set of  requirements for causality
are satisfied (e.g. $\alpha=4/3$, so that $\chi_{\{1,2\}}=3\chi_{\{3,4\}}=\lambda$) but there is no geometric reason at this
order we can demand to argue for a specific value.

\section{Final words}\label{sec:final}

In this work we examined from a gravitation angle how the BDNK first order 
formulation of relativistic, viscous hydrodynamics is connected to the fluid-gravity correspondence.
Such a formulation, which in practice is simpler to deal with than standard, second order
viscous formulations~\cite{Muller:1967zza,Israel:1979wp}, has received significant 
attention in recent years both at the theoretical
level \cite{Bemfica:2019knx, Kovtun:2019hdm, Bemfica:2019hok, Hoult:2020eho, Bemfica:2020zjp} and also in incipient numerical investigations (e.g.~\cite{Pandya:2021ief,Danielsson:2021ykm,Bantilan:2022ech}). The results obtained 
also revealed new 
connections between
relativistic and Carrollian hydrodynamics as well
as with gravity.

Our analysis unearthed a natural way to motivate the BDNK formulation from a gravitational perspective.
Further, the expected functional dependence of transport coefficients was obtained and,
for the viscous and heat-flux coefficients, a unique expression was found. As well, our analysis
 revealed a connection between the effective Carrollian hydrodynamic description of 
null surfaces and the asymptotic relativistic fluid that is identified at the timelike infinity of
perturbed black branes in AdS.  Such connection implies that, at leading order, Raychaudhuri's
and Damour's equations encode the conservation of a conformal perfect fluid. The analysis of 
higher orders and the exploration of Carrollian hydrodynamics from this perspective 
is an interesting task which we defer to future work.

In a similar vein, it would be interesting to explore the  horizon physics deeper, as results there would also hold for asymptotically flat spacetimes. Importantly, in the latter case,
there is also an interesting relation between the structure
of interior null surfaces  (like the horizon), and future null
infinity. However,
 the relationship between the horizon membrane paradigm and the 
 asymptotic (e.g. ${\cal I}^+$) null boundary Carrollian fluid is still
largely unexplored. The latter fluid however enjoys Weyl symmetry, which makes it special. This could also help motivate a
fluid interpretation of (particular) quasi-normal modes in asymptotically flat spacetimes. Another avenue for exploration
is to consider a potential entropy current, both for the relativistic fluid at the boundary and the horizon Carrollian fluid. This current  could help us connect with its microscopic origin and inform
standing questions on Carrolllian hydrodynamics.
Finally, a deeper understanding of potential connections between phenomena
in non-linear gravity and hydrodynamics can motivate new avenues to 
identify and study non-linear gravitational behavior (e.g.~\cite{VanRaamsdonk:2008fp,Eling:2010vr,Yang:2014tla,Adams:2013vsa,Carrasco:2012nf,Green:2013zba,Westernacher-Schneider:2017xie,Aretakis:2018dzy,Moschidis:2018mxs,Redondo-Yuste:2022czg}).

\acknowledgements
We thank F. Bemfica, M. Disconzi, L. Freidel, S. Giri, R. E. Hoult, 
P. Kovtun, R. Leigh, J. Noronha, M. Petropoulos, 
E. Poisson, F. Pretorius, M. Rangamani, and J. Senovilla
for discussions.
This work was supported in
part by Perimeter Institute for Theoretical Physics.
Research at Perimeter Institute is supported by the
Government of Canada through the Department of
Innovation, Science and Economic Development Canada
and by the Province of Ontario through the Ministry of
Economic Development, Job Creation and Trade. L. L.
thanks financial support via the Carlo Fidani Rainer Weiss
Chair at Perimeter Institute. L. L. receives additional
financial support from the Natural Sciences and
Engineering Research Council of Canada through a
Discovery Grant and CIFAR.
L. L. thanks KITP - UC Santa Barbara for its
hospitality during “The Many Faces of Relativistic Fluid Dynamics” Program, where this work’s initial stages were completed.

\subsection{Alternative choice for $\chi$}\label{altchoice}
An alternative choice for $\chi$, which fixes it completely,
would be to demand that to first order $\kappa=2/b$. This would
imply
\beqn
\chi &=& 2 \left( \frac{2b}{3} \partial_a u^a - u^a \partial_a b \right) \nonumber \\
&=& \frac{2b}{3} \left( 2 \partial_a u^a +  \frac{u^a \partial_a e}{(e+p)}\right),
\eeqn
and as a consequence,
$\chi_1=3\chi_3=2/b^3$, $\chi_2=3\chi_4=4/b^3$.
These values however, (complemented
by $\lambda=4/b^3,\eta=1/b^3$) are not
within the causality bounds of
~\cite{Bemfica:2019knx}. Further, on-shell dynamical solutions
have an associated energy density at first order which differs
from that at zeroth order. 

Going further, one could demand that the first order
expression for $\mu$ be proportional to $\Theta$, so first order perturbations ($\Theta \rightarrow \Theta + \delta \Theta$)  of Raychauduri's equation would receive no corrections from the non-linear terms except the shear contribution. 
In turn,
this would require $\kappa = 2/b + \alpha \Theta$, thus
\begin{equation}
\chi = \frac{b}{3} \left( (4+6\alpha) \partial_a u^a + (2+6\alpha) \frac{u^a \partial_a e}{(e+p)}
\right) \, .
\end{equation}
Thus, $\chi_1=3\chi_3=(2+ 6\alpha)/b^3$, $\chi_2=3\chi_4=(4+6\alpha)/b^3$.
For $\alpha\ge1$ the causality conditions are satisfied,
but again the associated energy-density at first order --on-shell-- would differ from that at zeroth
order.

\bibliographystyle{uiuchept}
\bibliography{arxivV2}

\providecommand{\href}[2]{#2}\begingroup\raggedright\begin{thebibliography}{10}

\bibitem{Bemfica:2017wps}
F.~S. Bemfica, M.~M. Disconzi, and J.~Noronha, ``{Causality and existence of
  solutions of relativistic viscous fluid dynamics with gravity},''
  \href{http://dx.doi.org/10.1103/PhysRevD.98.104064}{{\em Phys. Rev. D} {\bf
  98} (2018) no.~10, 104064}, \href{http://arxiv.org/abs/1708.06255}{{\tt
  arXiv:1708.06255 [gr-qc]}}.

\bibitem{Kovtun:2019hdm}
P.~Kovtun, ``{First-order relativistic hydrodynamics is stable},''
  \href{http://dx.doi.org/10.1007/JHEP10(2019)034}{{\em JHEP} {\bf 10} (2019)
  034}, \href{http://arxiv.org/abs/1907.08191}{{\tt arXiv:1907.08191
  [hep-th]}}.

\bibitem{Bemfica:2019knx}
F.~S. Bemfica, M.~M. Disconzi, and J.~Noronha, ``{Nonlinear Causality of
  General First-Order Relativistic Viscous Hydrodynamics},''
  \href{http://dx.doi.org/10.1103/PhysRevD.100.104020}{{\em Phys. Rev. D} {\bf
  100} (2019) no.~10, 104020}, \href{http://arxiv.org/abs/1907.12695}{{\tt
  arXiv:1907.12695 [gr-qc]}}. [Erratum: Phys.Rev.D 105, 069902 (2022)].

\bibitem{Hoult:2020eho}
R.~E. Hoult and P.~Kovtun, ``{Stable and causal relativistic Navier-Stokes
  equations},'' \href{http://dx.doi.org/10.1007/JHEP06(2020)067}{{\em JHEP}
  {\bf 06} (2020)  067}, \href{http://arxiv.org/abs/2004.04102}{{\tt
  arXiv:2004.04102 [hep-th]}}.

\bibitem{Bemfica:2020zjp}
F.~S. Bemfica, M.~M. Disconzi, and J.~Noronha, ``{First-Order
  General-Relativistic Viscous Fluid Dynamics},''
  \href{http://dx.doi.org/10.1103/PhysRevX.12.021044}{{\em Phys. Rev. X} {\bf
  12} (2022) no.~2, 021044}, \href{http://arxiv.org/abs/2009.11388}{{\tt
  arXiv:2009.11388 [gr-qc]}}.

\bibitem{Hoult:2021gnb}
R.~E. Hoult and P.~Kovtun, ``{Causal first-order hydrodynamics from kinetic
  theory and holography},''
  \href{http://dx.doi.org/10.1103/PhysRevD.106.066023}{{\em Phys. Rev. D} {\bf
  106} (2022) no.~6, 066023}, \href{http://arxiv.org/abs/2112.14042}{{\tt
  arXiv:2112.14042 [hep-th]}}.

\bibitem{GEROCH1991394}
R.~Geroch and L.~Lindblom, ``Causal theories of dissipative relativistic
  fluids,''
  \href{http://dx.doi.org/https://doi.org/10.1016/0003-4916(91)90063-E}{{\em
  Annals of Physics} {\bf 207} (1991) no.~2, 394--416}.

\bibitem{Denicol:2008ha}
G.~S. Denicol, T.~Kodama, T.~Koide, and P.~Mota, ``{Stability and Causality in
  relativistic dissipative hydrodynamics},''
  \href{http://dx.doi.org/10.1088/0954-3899/35/11/115102}{{\em J. Phys. G} {\bf
  35} (2008)  115102}, \href{http://arxiv.org/abs/0807.3120}{{\tt
  arXiv:0807.3120 [hep-ph]}}.

\bibitem{Pu:2009fj}
S.~Pu, T.~Koide, and D.~H. Rischke, ``{Does stability of relativistic
  dissipative fluid dynamics imply causality?},''
  \href{http://dx.doi.org/10.1103/PhysRevD.81.114039}{{\em Phys. Rev. D} {\bf
  81} (2010)  114039}, \href{http://arxiv.org/abs/0907.3906}{{\tt
  arXiv:0907.3906 [hep-ph]}}.

\bibitem{Lehner:2017yes}
L.~Lehner, O.~A. Reula, and M.~E. Rubio, ``{Hyperbolic theory of relativistic
  conformal dissipative fluids},''
  \href{http://dx.doi.org/10.1103/PhysRevD.97.024013}{{\em Phys. Rev. D} {\bf
  97} (2018) no.~2, 024013}, \href{http://arxiv.org/abs/1710.08033}{{\tt
  arXiv:1710.08033 [gr-qc]}}.

\bibitem{Aguilar:2017ios}
M.~Aguilar and E.~Calzetta, ``{Causal Relativistic Hydrodynamics of Conformal
  Fermi-Dirac Gases},''
  \href{http://dx.doi.org/10.1103/PhysRevD.95.076022}{{\em Phys. Rev. D} {\bf
  95} (2017) no.~7, 076022}, \href{http://arxiv.org/abs/1701.01916}{{\tt
  arXiv:1701.01916 [hep-ph]}}.

\bibitem{1959flme}
L.~D. {Landau} and E.~M. {Lifshitz}, {\em {Fluid mechanics}}.
\newblock 1959.

\bibitem{Eckart:1940te}
C.~Eckart, ``{The Thermodynamics of irreversible processes. 3.. Relativistic
  theory of the simple fluid},''
  \href{http://dx.doi.org/10.1103/PhysRev.58.919}{{\em Phys. Rev.} {\bf 58}
  (1940)  919--924}.

\bibitem{Bhattacharyya:2007vjd}
S.~Bhattacharyya, V.~E. Hubeny, S.~Minwalla, and M.~Rangamani, ``{Nonlinear
  Fluid Dynamics from Gravity},''
  \href{http://dx.doi.org/10.1088/1126-6708/2008/02/045}{{\em JHEP} {\bf 02}
  (2008)  045}, \href{http://arxiv.org/abs/0712.2456}{{\tt arXiv:0712.2456
  [hep-th]}}.

\bibitem{Bhattacharyya:2008xc}
S.~Bhattacharyya, V.~E. Hubeny, R.~Loganayagam, G.~Mandal, S.~Minwalla,
  T.~Morita, M.~Rangamani, and H.~S. Reall, ``{Local Fluid Dynamical Entropy
  from Gravity},'' \href{http://dx.doi.org/10.1088/1126-6708/2008/06/055}{{\em
  JHEP} {\bf 06} (2008)  055}, \href{http://arxiv.org/abs/0803.2526}{{\tt
  arXiv:0803.2526 [hep-th]}}.

\bibitem{VanRaamsdonk:2008fp}
M.~Van~Raamsdonk, ``{Black Hole Dynamics From Atmospheric Science},''
  \href{http://dx.doi.org/10.1088/1126-6708/2008/05/106}{{\em JHEP} {\bf 05}
  (2008)  106}, \href{http://arxiv.org/abs/0802.3224}{{\tt arXiv:0802.3224
  [hep-th]}}.

\bibitem{Haack:2008cp}
M.~Haack and A.~Yarom, ``{Nonlinear viscous hydrodynamics in various dimensions
  using AdS/CFT},'' \href{http://dx.doi.org/10.1088/1126-6708/2008/10/063}{{\em
  JHEP} {\bf 10} (2008)  063}, \href{http://arxiv.org/abs/0806.4602}{{\tt
  arXiv:0806.4602 [hep-th]}}.

\bibitem{Banerjee:2008th}
N.~Banerjee, J.~Bhattacharya, S.~Bhattacharyya, S.~Dutta, R.~Loganayagam, and
  P.~Surowka, ``{Hydrodynamics from charged black branes},''
  \href{http://dx.doi.org/10.1007/JHEP01(2011)094}{{\em JHEP} {\bf 01} (2011)
  094}, \href{http://arxiv.org/abs/0809.2596}{{\tt arXiv:0809.2596 [hep-th]}}.

\bibitem{Rangamani:2009xk}
M.~Rangamani, ``{Gravity and Hydrodynamics: Lectures on the fluid-gravity
  correspondence},''
  \href{http://dx.doi.org/10.1088/0264-9381/26/22/224003}{{\em Class. Quant.
  Grav.} {\bf 26} (2009)  224003}, \href{http://arxiv.org/abs/0905.4352}{{\tt
  arXiv:0905.4352 [hep-th]}}.

\bibitem{Caldarelli:2011idw}
M.~M. Caldarelli, R.~G. Leigh, A.~C. Petkou, P.~M. Petropoulos, V.~Pozzoli, and
  K.~Siampos, ``{Vorticity in holographic fluids},''
  \href{http://dx.doi.org/10.22323/1.155.0076}{{\em PoS} {\bf CORFU2011} (2011)
   076}, \href{http://arxiv.org/abs/1206.4351}{{\tt arXiv:1206.4351 [hep-th]}}.

\bibitem{Mukhopadhyay:2013gja}
A.~Mukhopadhyay, A.~C. Petkou, P.~M. Petropoulos, V.~Pozzoli, and K.~Siampos,
  ``{Holographic perfect fluidity, Cotton energy-momentum duality and transport
  properties},'' \href{http://dx.doi.org/10.1007/JHEP04(2014)136}{{\em JHEP}
  {\bf 04} (2014)  136}, \href{http://arxiv.org/abs/1309.2310}{{\tt
  arXiv:1309.2310 [hep-th]}}.

\bibitem{Gath:2015nxa}
J.~Gath, A.~Mukhopadhyay, A.~C. Petkou, P.~M. Petropoulos, and K.~Siampos,
  ``{Petrov Classification and holographic reconstruction of spacetime},''
  \href{http://dx.doi.org/10.1007/JHEP09(2015)005}{{\em JHEP} {\bf 09} (2015)
  005}, \href{http://arxiv.org/abs/1506.04813}{{\tt arXiv:1506.04813
  [hep-th]}}.

\bibitem{Damour:1979wya}
T.~Damour, {\em {Quelques proprietes mecaniques, electromagnetiques,
  thermodynamiques et quantiques des trous noir}}.
\newblock PhD thesis, Paris U., VI-VII, 1979.

\bibitem{Thorne:1986iy}
K.~S. Thorne, R.~H. Price, and D.~A. Macdonald, eds., {\em {Black holes: the
  membrane paradigm}}.
\newblock 1986.

\bibitem{Penna:2018gfx}
R.~F. Penna, ``{Near-horizon Carroll symmetry and black hole Love numbers},''
  \href{http://arxiv.org/abs/1812.05643}{{\tt arXiv:1812.05643 [hep-th]}}.

\bibitem{Donnay:2019jiz}
L.~Donnay and C.~Marteau, ``{Carrollian Physics at the Black Hole Horizon},''
  \href{http://dx.doi.org/10.1088/1361-6382/ab2fd5}{{\em Class. Quant. Grav.}
  {\bf 36} (2019) no.~16, 165002}, \href{http://arxiv.org/abs/1903.09654}{{\tt
  arXiv:1903.09654 [hep-th]}}.

\bibitem{Ciambelli:2019lap}
L.~Ciambelli, R.~G. Leigh, C.~Marteau, and P.~M. Petropoulos, ``{Carroll
  Structures, Null Geometry and Conformal Isometries},''
  \href{http://dx.doi.org/10.1103/PhysRevD.100.046010}{{\em Phys. Rev. D} {\bf
  100} (2019) no.~4, 046010}, \href{http://arxiv.org/abs/1905.02221}{{\tt
  arXiv:1905.02221 [hep-th]}}.

\bibitem{Freidel:2022bai}
L.~Freidel and P.~Jai-akson, ``{Carrollian hydrodynamics from symmetries},''
  \href{http://dx.doi.org/10.1088/1361-6382/acb194}{{\em Class. Quant. Grav.}
  {\bf 40} (2023) no.~5, 055009}, \href{http://arxiv.org/abs/2209.03328}{{\tt
  arXiv:2209.03328 [hep-th]}}.

\bibitem{Freidel:2022vjq}
L.~Freidel and P.~Jai-akson, ``{Carrollian hydrodynamics and symplectic
  structure on stretched horizons},''
  \href{http://arxiv.org/abs/2211.06415}{{\tt arXiv:2211.06415 [gr-qc]}}.

\bibitem{Redondo-Yuste:2022czg}
J.~Redondo-Yuste and L.~Lehner, ``{Non-linear black hole dynamics and
  Carrollian fluids},'' \href{http://dx.doi.org/10.1007/JHEP02(2023)240}{{\em
  JHEP} {\bf 02} (2023)  240}, \href{http://arxiv.org/abs/2212.06175}{{\tt
  arXiv:2212.06175 [gr-qc]}}.

\bibitem{Ciambelli:2023mir}
L.~Ciambelli, L.~Freidel, and R.~G. Leigh, ``{Null Raychaudhuri: Canonical
  Structure and the Dressing Time},''
  \href{http://arxiv.org/abs/2309.03932}{{\tt arXiv:2309.03932 [hep-th]}}.

\bibitem{Henningson:1998gx}
M.~Henningson and K.~Skenderis, ``{The Holographic Weyl anomaly},''
  \href{http://dx.doi.org/10.1088/1126-6708/1998/07/023}{{\em JHEP} {\bf 07}
  (1998)  023}, \href{http://arxiv.org/abs/hep-th/9806087}{{\tt
  arXiv:hep-th/9806087}}.

\bibitem{Balasubramanian:1999re}
V.~Balasubramanian and P.~Kraus, ``{A Stress tensor for Anti-de Sitter
  gravity},'' \href{http://dx.doi.org/10.1007/s002200050764}{{\em Commun. Math.
  Phys.} {\bf 208} (1999)  413--428},
  \href{http://arxiv.org/abs/hep-th/9902121}{{\tt arXiv:hep-th/9902121}}.

\bibitem{Mars:1993mj}
M.~Mars and J.~M.~M. Senovilla, ``{Geometry of general hypersurfaces in
  space-time: Junction conditions},''
  \href{http://dx.doi.org/10.1088/0264-9381/10/9/026}{{\em Class. Quant. Grav.}
  {\bf 10} (1993)  1865--1897}, \href{http://arxiv.org/abs/gr-qc/0201054}{{\tt
  arXiv:gr-qc/0201054}}.

\bibitem{Berti:2009kk}
E.~Berti, V.~Cardoso, and A.~O. Starinets, ``{Quasinormal modes of black holes
  and black branes},''
  \href{http://dx.doi.org/10.1088/0264-9381/26/16/163001}{{\em Class. Quant.
  Grav.} {\bf 26} (2009)  163001}, \href{http://arxiv.org/abs/0905.2975}{{\tt
  arXiv:0905.2975 [gr-qc]}}.

\bibitem{Muller:1967zza}
I.~Muller, ``{Zum Paradoxon der Warmeleitungstheorie},''
  \href{http://dx.doi.org/10.1007/BF01326412}{{\em Z. Phys.} {\bf 198} (1967)
  329--344}.

\bibitem{Israel:1979wp}
W.~Israel and J.~M. Stewart, ``{Transient relativistic thermodynamics and
  kinetic theory},'' \href{http://dx.doi.org/10.1016/0003-4916(79)90130-1}{{\em
  Annals Phys.} {\bf 118} (1979)  341--372}.

\bibitem{Bemfica:2019hok}
F.~S. Bemfica, M.~M. Disconzi, C.~Rodriguez, and Y.~Shao, ``{Local
  well-posedness in Sobolev spaces for first-order conformal causal
  relativistic viscous hydrodynamics},''
  \href{http://arxiv.org/abs/1911.02504}{{\tt arXiv:1911.02504 [math.AP]}}.

\bibitem{Pandya:2021ief}
A.~Pandya and F.~Pretorius, ``{Numerical exploration of first-order
  relativistic hydrodynamics},''
  \href{http://dx.doi.org/10.1103/PhysRevD.104.023015}{{\em Phys. Rev. D} {\bf
  104} (2021) no.~2, 023015}, \href{http://arxiv.org/abs/2104.00804}{{\tt
  arXiv:2104.00804 [gr-qc]}}.

\bibitem{Danielsson:2021ykm}
U.~Danielsson, L.~Lehner, and F.~Pretorius, ``{Dynamics and observational
  signatures of shell-like black hole mimickers},''
  \href{http://dx.doi.org/10.1103/PhysRevD.104.124011}{{\em Phys. Rev. D} {\bf
  104} (2021) no.~12, 124011}, \href{http://arxiv.org/abs/2109.09814}{{\tt
  arXiv:2109.09814 [gr-qc]}}.

\bibitem{Bantilan:2022ech}
H.~Bantilan, Y.~Bea, and P.~Figueras, ``{Evolutions in first-order viscous
  hydrodynamics},'' \href{http://dx.doi.org/10.1007/JHEP08(2022)298}{{\em JHEP}
  {\bf 08} (2022)  298}, \href{http://arxiv.org/abs/2201.13359}{{\tt
  arXiv:2201.13359 [hep-th]}}.

\bibitem{Eling:2010vr}
C.~Eling, I.~Fouxon, and Y.~Oz, ``{Gravity and a Geometrization of Turbulence:
  An Intriguing Correspondence},'' \href{http://arxiv.org/abs/1004.2632}{{\tt
  arXiv:1004.2632 [hep-th]}}.

\bibitem{Yang:2014tla}
H.~Yang, A.~Zimmerman, and L.~Lehner, ``{Turbulent Black Holes},''
  \href{http://dx.doi.org/10.1103/PhysRevLett.114.081101}{{\em Phys. Rev.
  Lett.} {\bf 114} (2015)  081101}, \href{http://arxiv.org/abs/1402.4859}{{\tt
  arXiv:1402.4859 [gr-qc]}}.

\bibitem{Adams:2013vsa}
A.~Adams, P.~M. Chesler, and H.~Liu, ``{Holographic turbulence},''
  \href{http://dx.doi.org/10.1103/PhysRevLett.112.151602}{{\em Phys. Rev.
  Lett.} {\bf 112} (2014) no.~15, 151602},
  \href{http://arxiv.org/abs/1307.7267}{{\tt arXiv:1307.7267 [hep-th]}}.

\bibitem{Carrasco:2012nf}
F.~Carrasco, L.~Lehner, R.~C. Myers, O.~Reula, and A.~Singh, ``{Turbulent flows
  for relativistic conformal fluids in 2+1 dimensions},''
  \href{http://dx.doi.org/10.1103/PhysRevD.86.126006}{{\em Phys. Rev. D} {\bf
  86} (2012)  126006}, \href{http://arxiv.org/abs/1210.6702}{{\tt
  arXiv:1210.6702 [hep-th]}}.

\bibitem{Green:2013zba}
S.~R. Green, F.~Carrasco, and L.~Lehner, ``{Holographic Path to the Turbulent
  Side of Gravity},'' \href{http://dx.doi.org/10.1103/PhysRevX.4.011001}{{\em
  Phys. Rev. X} {\bf 4} (2014) no.~1, 011001},
  \href{http://arxiv.org/abs/1309.7940}{{\tt arXiv:1309.7940 [hep-th]}}.

\bibitem{Westernacher-Schneider:2017xie}
J.~R. Westernacher-Schneider, ``{Fractal dimension of turbulent black holes},''
  \href{http://dx.doi.org/10.1103/PhysRevD.96.104054}{{\em Phys. Rev. D} {\bf
  96} (2017) no.~10, 104054}, \href{http://arxiv.org/abs/1710.04264}{{\tt
  arXiv:1710.04264 [gr-qc]}}.

\bibitem{Aretakis:2018dzy}
S.~Aretakis, \href{http://dx.doi.org/10.1007/978-3-319-95183-6}{{\em {Dynamics
  of Extremal Black Holes}}}, vol.~33 of {\em SpringerBriefs in Mathematical
  Physics}.
\newblock Springer International Publishing, Cham, 2018.

\bibitem{Moschidis:2018mxs}
G.~Moschidis, {\em {Two instability results in general relativity}}.
\newblock PhD thesis, Princeton U., Math. Dept., 2018.

\end{thebibliography}\endgroup

\end{document}